\documentclass{article}
\usepackage{iclr2027_conference,times}
\iclrpreprintcopy
\usepackage{amsmath,amsfonts,bm}

\def\eqref#1{equation~\ref{#1}}
\def\1{\bm{1}}

\DeclareMathAlphabet{\mathsfit}{\encodingdefault}{\sfdefault}{m}{sl}
\SetMathAlphabet{\mathsfit}{bold}{\encodingdefault}{\sfdefault}{bx}{n}

\usepackage{graphicx,booktabs,tabularx,tikz}
\usetikzlibrary{arrows.meta,positioning,fit}
\PassOptionsToPackage{hyphens}{url}
\usepackage{hyperref}
\hypersetup{hidelinks,
  pdftitle={ECO: Energy-Oriented Configuration Optimization for Attention--FFN Disaggregated LLM Serving},
  pdfauthor={Zou Qingyun, Bin Gao, Zhuobin Huang, Weng-Fai Wong, Tulika Mitra, Bingsheng He}}
\AtBeginDocument{%
  \setlength{\abovedisplayskip}{4pt plus 1pt minus 1pt}%
  \setlength{\belowdisplayskip}{4pt plus 1pt minus 1pt}%
  \setlength{\abovedisplayshortskip}{0pt plus 1pt}%
  \setlength{\belowdisplayshortskip}{2pt plus 1pt minus 1pt}%
}
\makeatletter
\def\@textbottom{\ifnum\c@page>5\ifnum\c@page<8\vskip\z@\@plus .0001fil\relax\fi\fi}
\makeatother
\newcommand{\ctwobo}{\texorpdfstring{C\textsuperscript{2}BO}{C2BO}}

\title{ECO: Energy-Oriented Configuration Optimization for Attention–FFN Disaggregated LLM Serving}
\author{\textbf{Zou Qingyun, Bin Gao, Zhuobin Huang,}\\
\textbf{Weng-Fai Wong, Tulika Mitra, Bingsheng He}\\
\normalfont National University of Singapore}
\begin{document}
\maketitle
\begin{abstract}
Energy-efficient LLM serving requires minimizing serving GPU energy while meeting latency and throughput service-level objectives (SLOs).
Attention--FFN disaggregation (AFD) enables separate resource allocation and operating controls for attention and expert computation, but their energy effects remain coupled through the execution pipeline.
Realizing its energy-saving potential therefore requires navigating a hierarchical configuration space in which deployment structures constrain admissible controls and shape their end-to-end effects.
Finding low-energy configurations that meet SLOs is challenging because physical evaluations are costly and only a small fraction of candidates can be measured.
We present Energy-Oriented Configuration Optimization (ECO), which jointly searches deployment structures and their admissible operating controls under a limited measurement budget.
ECO constructs a structure-aware energy prior from calibrated stage behavior and pipeline dependencies, then learns residual prediction errors with a Gaussian process.
Its cost-aware constrained Bayesian optimization prioritizes measurements according to expected energy improvement while accounting for SLO feasibility, execution success, and evaluation cost, and returns the lowest-energy measured feasible configuration.
Across all 16 scenarios on A6000 and A100 with Qwen and DeepSeek, ECO's frozen configurations, evaluated on disjoint requests, reduce serving energy by 40.5\% and increase output token rate by 20.7\% on average relative to baselines while meeting target SLOs.
Across the 8 A6000 scenarios, its selected feasible energy averages 33.1\% below generic constrained Bayesian optimization and 25.8\% below genetic search.
\end{abstract}
\suppressfloats[t]
\section{Introduction}
\label{sec:introduction}

Energy-efficient LLM serving requires minimizing serving GPU energy while meeting latency and throughput service-level objectives (SLOs).
Attention--feed-forward network (FFN) disaggregation (AFD) provides additional flexibility for pursuing this objective.
Attention and expert FFNs have different computation, memory, and communication demands~\citep{rajbhandari2022deepspeedmoe,zhu2025megascale}.
By placing them on separate GPU groups, AFD allows each role to use its own resource allocation, parallelism, frequency, and power cap~\citep{zhu2025megascale,wu2026disaggregation}.
By placing them on separate GPU groups, AFD provides flexibility in both the \emph{deployment structure}, including GPU allocation, parallelism, and dual batch overlap (DBO) settings, and the \emph{operating controls}, including per-role frequencies and power caps~\citep{zhu2025megascale,wu2026disaggregation}.
The central configuration problem is how to use this flexibility to reduce energy across the active GPUs while satisfying fixed service requirements.
Three properties of AFD make this problem challenging.

\begin{figure}[t]
  \vspace{-6pt}
  \centering
  \includegraphics[width=\linewidth]{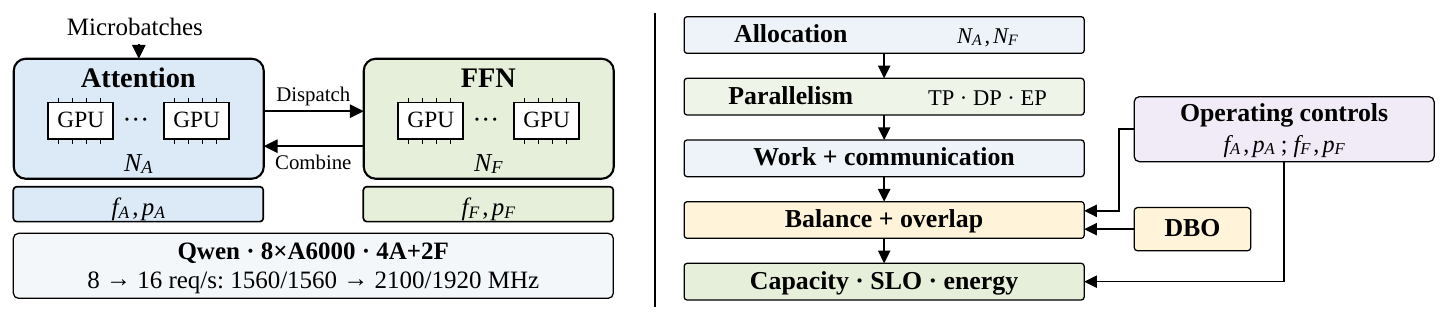}
  \vspace{-15pt}
  \caption{AFD execution and configuration dependencies. Left: GPU groups and load-dependent clocks. Right: configuration choices and their effects on service and energy.}
  \label{fig:afd-motivation}
  \vspace{-15pt}
\end{figure}

\textit{First}, deployment structure and operating controls are coupled through the execution pipeline.
Each microbatch passes through attention, dispatch, FFN, and combine before entering the next layer.
Slowing one stage may reduce its power draw, but can also delay dependent stages and extend the time during which other GPUs consume power.
A local power reduction therefore need not reduce total serving energy: its benefit depends on how the selected structure permits work to overlap and delays to propagate.
Evaluating operating controls therefore requires accounting for their effects across the pipeline.

\textit{Second}, AFD has a hierarchical configuration space.
GPU allocation constrains the parallelism choices available to each role, while parallelism and DBO settings determine how work is partitioned and overlapped within the deployment.
Frequencies and power caps must also satisfy the constraints of the assigned hardware.
These dependencies restrict which combinations of settings form valid configurations and must be respected during search.
However, satisfying the known static constraints does not ensure that a configuration will execute successfully or meet the SLOs.

\textit{Third}, configuration selection depends on the model and offered load.
The same deployment structure and operating controls can produce different energy and service responses across models and workloads.
As load increases, a configuration may need higher operating frequencies, additional GPUs, or a different allocation between attention and FFN computation to remain feasible.
The appropriate adjustment can differ across models, so deployment structures and operating controls need to be evaluated jointly for the target model and workload.

Identifying suitable configurations under these conditions requires extensive measurements, but each measurement is costly.
Each trial incurs lifecycle overhead from deployment, warmup, workload replay, and telemetry collection, and failed trials consume budget too.
Our audited A6000 spaces contain 2,700--22,950 statically legal candidates, while 16 search attempts cover only 0.070--0.593\%.
Such sparse coverage makes it difficult to learn the energy and feasibility of unmeasured configurations solely from observed outcomes.
The practical problem is therefore to identify low-energy, SLO-feasible configurations in a structured search space using a limited measurement budget.

Prior systems establish the value of coordinated provisioning and frequency control.
AFD systems study resource partitioning and communication-aware execution~\citep{zhu2025megascale,wu2026disaggregation}; DynamoLLM coordinates instance count, parallelism, and frequency~\citep{stojkovic2024dynamollm}; and AFlex combines AFD provisioning with operator profiling and GPU frequency control~\citep{hu2026aflex}.
Our focus is how to select energy-efficient, SLO-feasible configurations when measurements for the target deployment are limited.
A black-box energy surrogate without pipeline information must learn configuration effects primarily from sparse observations.
AFD, however, exposes useful structure before search begins: configuration rules constrain legal deployments, and pipeline dependencies shape how operating controls affect energy.
The key question is how to incorporate this information into an energy prior that helps identify promising feasible configurations within the available measurement budget.

We present \textbf{Energy-Oriented Configuration Optimization (ECO)} to address this question.
To our knowledge, ECO is the first design-space exploration framework jointly searching AFD structures and GPU frequencies under a limited measurement budget.
As search proceeds, stage records from measured configurations calibrate structure-specific stage behavior.
ECO propagates predicted stage times through pipeline dependencies and combines the resulting timing estimates with calibrated GPU power to estimate serving energy.
A Gaussian process (GP) learns residual prediction errors from completed trials, refining the prior with measured outcomes.
Building on constrained and cost-aware Bayesian optimization~\citep{gelbart2014constraints,gardner2014constraints,snoek2012practical}, ECO's cost-aware constrained Bayesian optimization (\ctwobo{}) prioritizes candidates according to expected energy improvement while accounting for SLO feasibility, execution success, and evaluation cost.
Acquisition-based choices are interleaved with structure and single-control probes that supply calibration data within the same search allowance.
ECO returns the lowest-energy measured configuration that passes the execution and SLO eligibility checks, with its settings fixed during subsequent serving.
Our contributions are as follows:

\begingroup
\setlength{\leftmargini}{0pt}
\begin{itemize}
    \item \textbf{Configuration characterization.}
    We characterize AFD's deployment hierarchy, present an example of structure-dependent frequency behavior, and examine model- and load-dependent configuration selections. Exact validator counts quantify search-space coverage under measurement budget.

    \item \textbf{Structure-guided optimization.}
    We develop ECO for budgeted AFD configuration search.
    ECO combines a calibrated stage-and-pipeline energy prior with GP residual correction and cost-aware constrained Bayesian optimization to guide the selection of physical measurements.

    \item \textbf{Empirical evaluation.}
    We evaluate ECO with Qwen and DeepSeek on A6000 and A100 GPUs.
    Across all 16 scenarios, the frozen configurations, evaluated on disjoint requests, reduce serving energy by 40.5\% and increase output token rate by 20.7\% on average relative to Default on the same held-out requests, with scenarios weighted equally, while meeting the fixed scenario SLOs.
    Ablations examine the effects of energy modeling, joint tuning, and cost-aware acquisition.

    \item \textbf{Deployment insights.}
    We examine the selected configurations to derive deployment insights into configuration reuse, asymmetric GPU allocation, and service headroom in the tested scenarios.
\end{itemize}
\endgroup

\section{AFD Configuration Landscape}
\label{sec:background}

AFD places attention and expert FFN computation on separate GPU groups.
Microbatches follow the attention--dispatch--FFN--combine pipeline across layers and can overlap when resources permit (Figure~\ref{fig:afd-motivation}).
This section examines how deployment structures and operating controls interact through the pipeline, how configuration choices depend on one another, and how model and load affect configuration selection under fixed SLOs.

\begin{figure}[t]
  \centering
  \includegraphics[width=\linewidth]{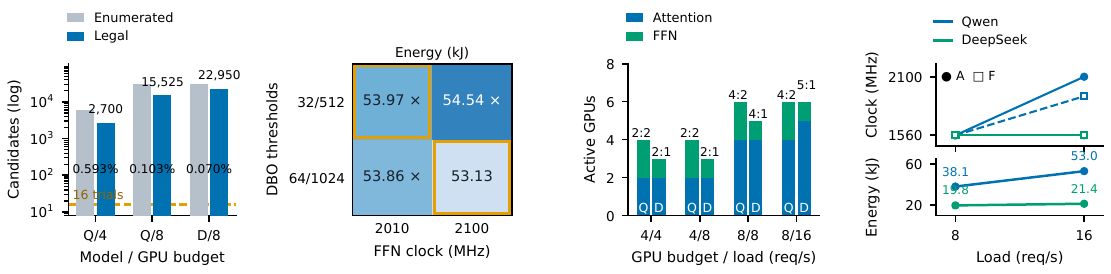}
  \vspace{-10pt}
  \caption{AFD configuration characterization on A6000. From left to right: search-space size, DBO--frequency interaction, selected allocations, and selected clocks and energy across loads.}
  \label{fig:afd-characterization}
  \vspace{-10pt}
\end{figure}

\textbf{Motivation 1: Structure--control coupling.}
A role's frequency affects both its power and the timing of dependent stages.
Slowing attention can delay FFN execution, while slowing FFN can delay the next layer.
A local power reduction can therefore extend the time during which other GPUs consume power, so its effect on total serving energy depends on how delays propagate through the pipeline.
DBO thresholds control when two microbatches overlap, and overlap and shared resources shape this propagation~\citep{yu2022orca,agrawal2024sarathi}.
Consequently, isolated operator energy profiles do not capture the full serving response.
The matched Qwen measurements at 16 requests/s in Figure~\ref{fig:afd-characterization} illustrate this dependence.
With other settings fixed, raising the FFN clock target from 2010 to 2100\,MHz increases energy under DBO thresholds $(32,512)$ but decreases it under $(64,1024)$.
The difference between these two energy changes is $-1.303$\,kJ (Appendix~\ref{app:characterization}).
Each cell contains one observation, however, and three cells violate the throughput SLO; repeated measurements are needed to distinguish the interaction from run-to-run variation.
We use this local reversal as motivation for evaluating operating controls within their deployment structure.

\textbf{Motivation 2: Hierarchical configuration space.}
GPU allocation constrains the parallelism available to each role.
Let $N_A$ and $N_F$ denote the numbers of attention and FFN GPUs.
Tensor parallelism (TP) partitions computation, data parallelism (DP) distributes request batches, and expert parallelism (EP) distributes experts.
In our backend, $\mathrm{TP}_r$ must divide $N_r$,
$\mathrm{DP}_r=N_r/\mathrm{TP}_r$ for $r\in\{A,F\}$,
and $\mathrm{EP}_F=N_F$; allocation additionally requires
$N_A\geq N_F$ and $N_F\mid N_A$.
For example, 4A+2F admits attention TP degrees $\{1,2,4\}$ and FFN TP degrees $\{1,2\}$ before model and memory checks.
Energy optimization must therefore respect the dependencies that define valid deployment structures and evaluate operating controls within those structures.
Figure~\ref{fig:afd-characterization} reports the resulting search-space sizes.
Static checks leave 2,700--22,950 candidates across the audited A6000 settings, while 16 search attempts cover only 0.070--0.593\%.
These checks establish static legality; execution success and SLO feasibility still require measurement (Appendix~\ref{app:characterization}).
The resulting sparse coverage motivates using known configuration structure to guide the limited physical evaluations.

\textbf{Motivation 3: Model and load dependence.}
The two rightmost panels of Figure~\ref{fig:afd-characterization} compare the best feasible configurations found for Qwen (Q) and DeepSeek (D).
Under an eight-GPU budget, increasing load from 8 to 16 requests/s leaves Qwen's 4A+2F deployment structure and power caps unchanged, but raises attention/FFN clocks from 1560/1560 to 2100/1920\,MHz.
DeepSeek instead changes allocation from 4A+1F to 5A+1F while retaining 1560/1560\,MHz clocks.
These selections show different deployment structures and operating controls across models as load increases, motivating their joint evaluation for the target workload.
They are the best feasible configurations found under limited search budgets, rather than established global optima, and do not isolate the causal effect of individual configuration choices.

Together, these observations and configuration constraints motivate ECO's design:
pipeline coupling motivates a stage-and-pipeline energy prior; the configuration hierarchy motivates structure-conditioned search; model and load dependence motivates stage calibration on the target workload; and sparse physical coverage motivates combining the prior with residual correction as measurements become available.

% Declare the workflow early so it appears at the top of page 4.
\begin{figure}[t]
  \vspace{-6pt}
\centering
\includegraphics[width=\linewidth]{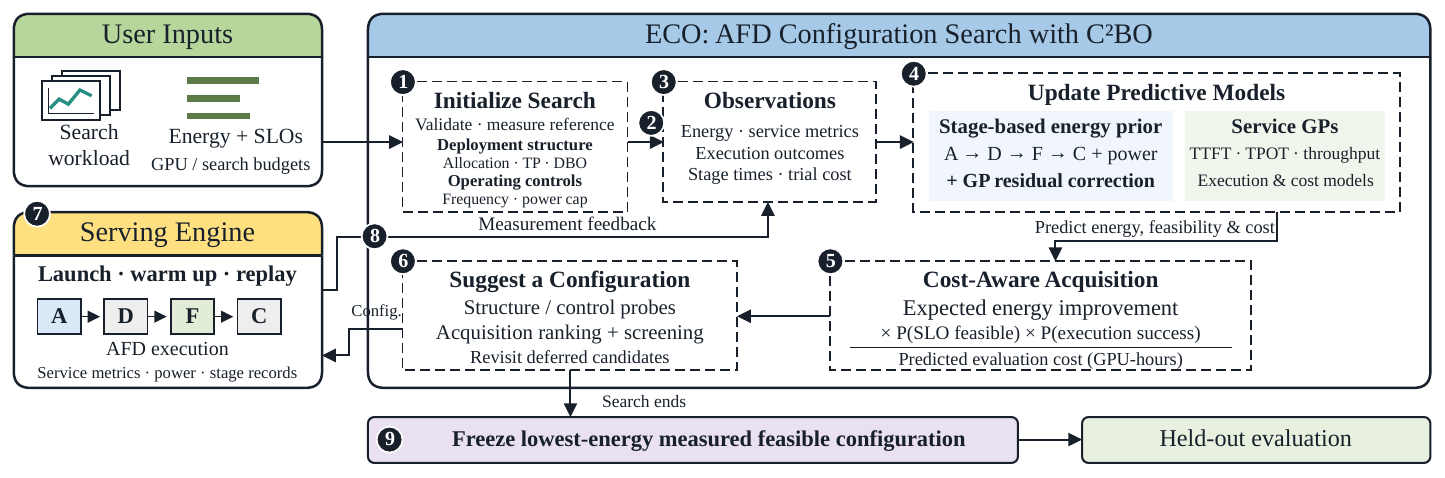}
\vspace{-10pt}
  \caption{ECO workflow. Stage profiling calibrates the energy prior; trial outcomes update predictive models. \ctwobo{} selects configurations on the fixed search workload. The final choice is frozen before independent held-out replay, which supplies the main results.}
\label{fig:eco-workflow}
  \vspace{-10pt}
\end{figure}

\section{Problem Formulation}
\label{sec:problem-formulation}

\textbf{Configuration space.} Given an LLM, a GPU budget $G$, a target workload $\mathcal W$, and SLO targets, we search for a static configuration $x=(s,u)$. The deployment structure $s=(N_A,N_F,\mathrm{TP}_A,\mathrm{TP}_F,b)$ specifies the GPU allocation and tensor parallelism of each role, together with a DBO setting $b$ that includes its enabled state and active decode/prefill thresholds. The operating controls $u=(f_A,f_F,p_A,p_F)$ specify per-role frequencies and power caps. DP and EP are derived from the structure under the backend's rules; placement and other serving settings remain fixed.

Let $\mathcal S$ contain deployment structures that satisfy known static constraints on allocation, divisibility, model compatibility, memory, hardware, and backend support, including $1\leq N_F\leq N_A$, $N_A+N_F\leq G$, and $N_F\mid N_A$. For each structure $s$, let $\mathcal U(s)$ contain operating controls permitted by its assigned hardware and static constraints. The search domain is $\mathcal X=\bigcup_{s\in\mathcal S}\bigl(\{s\}\times\mathcal U(s)\bigr)$. Structures may share the same operating-control grid while exhibiting different energy and service responses. Table~\ref{tab:design-space} specifies the experimental grids; DBO thresholds are active only when DBO is enabled. Membership in $\mathcal X$ establishes static legality, not execution success or SLO feasibility.

\textbf{Energy and feasibility.} Serving energy $E(x;\mathcal W)$ integrates power across the active GPUs from the first request submission to the last response completion, including waiting periods (Appendix~\ref{app:energy}). The feasible set $\mathcal F_{\mathcal W}\subseteq\mathcal X$ contains configurations that execute successfully, denoted by $v(x;\mathcal W)=1$, and meet all applicable service targets: p90 time to first token (TTFT) and time per output token (TPOT) must not exceed their thresholds, and output token rate must satisfy $R(x;\mathcal W)\geq R_{\min}$. Scenarios with a time-between-tokens (TTBT) constraint additionally require p90 TTBT to remain within its fixed threshold. We call these quantities service metrics and their SLO requirements service constraints. All thresholds remain unchanged throughout search. Energy, execution success, and service metrics are unknown for unmeasured configurations and must be estimated from measurements of the target workload.

\textbf{Measurement budget.} A search policy selects configuration $x_t$ using observations available before attempt $t$. Each attempt incurs cost $c_t$ in GPU-hours for deployment, warmup, replay with telemetry, and cleanup, including failed attempts. We allow $n\leq K$ search attempts and count initialization probes toward $K$. A shared preparation measurement is charged separately: the experiments use $K=16$ search attempts plus one preparation evaluation, for 17 charged evaluations in total. If preparation costs $C_{\mathrm{prep}}$ GPU-hours, the nominal time-budget constraint is $C_{\mathrm{prep}}+\sum_{t=1}^{n}c_t\leq B$. Trial cost depends on the preceding configuration and is uncertain before execution. ECO admits a trial only if an attempt remains and its predicted cost fits the remaining GPU-hour budget. If actual cost overruns this budget, ECO records the full cost and launches no further trials. Search quality is the lowest feasible energy physically evaluated on $\mathcal W$ within this accounting.

\textbf{Final selection.} Let $\mathcal M_n$ contain the measured configurations eligible under the execution and SLO checks, with repeated measurements aggregated as described in Appendix~\ref{app:eligibility}. Search returns
\begin{equation}
\hat x_n\in\operatorname*{arg\,min}_{x\in\mathcal M_n}\overline E_n(x;\mathcal W),
\label{eq:dse-selection}
\end{equation}
where $\overline E_n(x;\mathcal W)$ is the measured mean serving energy of configuration $x$ on the target workload. If $\mathcal M_n$ is empty, search reports that no feasible configuration was found. We freeze this selection and evaluate it on disjoint held-out requests. These replay measurements provide the main comparison between methods; they do not affect search or selection.

\section{ECO Methods}
\label{sec:eco-methods}

% \subsection{ECO Workflow}
% \label{sec:eco-workflow}
% \label{sec:eco-bo}

ECO calibrates structure-specific stage responses, propagates timing through the pipeline, and learns residual energy error (Figure~\ref{fig:eco-workflow}). After candidate validation and a reference measurement, \ctwobo{} updates its predictive models to select each trial. All trials use the same fixed search workload. We reserve calibration for fitting ECO's stage responses within the search budget. ECO returns the lowest-energy measured feasible configuration.

\subsection{Predictive Models}
\label{sec:eco-surrogates}
\label{sec:eco-prior}

The energy surrogate combines a calibrated stage-and-pipeline prior with GP residual correction. Calibration fits each deployment structure $s$ separately because stage responses to operating controls $u$ are structure dependent.

\textbf{Stage-time model.}
We decompose each layer into Attention ($A$), dispatch ($D$), FFN ($F$), and combine ($C$). Within each structure, nonnegative least squares fits stage duration to prefill and decode token counts. Paired measurements that vary one operating control estimate its effect on stage time. Multiple changes combine these local responses unless calibration identifies a joint effect; unidentified effects and extrapolation increase prior uncertainty. Appendix~\ref{app:surrogate-details} gives the fitting equations and implementation details.

\textbf{Pipeline and energy model.}
First-in, first-out (FIFO) simulator propagates predicted stage times through a finite number of in-flight microbatches, capturing execution time. An operation starts when its predecessor has completed and its resource is free: $F_o=\max\{F_{\mathrm{pred}(o)},A_{\mathrm{res}(o)}\}+\widehat s_o$, where $F_o$ is operation $o$'s finish time and $A_{\mathrm{res}(o)}$ is its resource's availability time. This recurrence propagates local timing changes through dependencies and resource contention. A microbatch enters the next layer only after combine, and two communication directions share resource by default. The completion time $\widehat T_{\mathrm{pipe}}$ gives the capacity estimate $\widehat C=N_{\mathrm{ref}}/\widehat T_{\mathrm{pipe}}$, where $N_{\mathrm{ref}}$ is the calibrated number of requests per pipeline unit. For $N$ requests at arrival rate $\lambda$, the log-energy prior is
\begin{equation}
% \small
 m_0(x)=\log\!\left(\frac{N\widehat P(x)}{\min\{\lambda,\widehat C(x)\}}\right).
 \label{eq:eco-stage-energy}
\end{equation}
Here $\widehat P(x)$ is calibrated total GPU power, bounded by the configured caps, and time is in seconds. The effective processing rate is limited by arrival rate or pipeline capacity. This approximation ranks candidates by energy; separate service models predict SLO feasibility.

\textbf{GP correction.}
The GP learns the prior's error from completed trials. With prior log-energy mean $m_0$ and uncertainty $s_0>0$, the residual target for successful measurement $i$ is $y_i=(\log E_i-m_0(x_i))/s_0(x_i)$. For residual posterior mean $\mu_h(x)$ and standard deviation $\sigma_h(x)$, the corrected prediction is $\mu_E(x)=m_0(x)+s_0(x)\mu_h(x)$ with $\sigma_E(x)=s_0(x)\sigma_h(x)$. Prior updates trigger recomputation of all residual targets before refitting. Appendix~\ref{app:surrogate-details} specifies GP inputs and kernel.

\textbf{Service and execution models.}
Separate GPs model log ratios of latency to its upper limit and the throughput floor to achieved throughput; nonpositive values satisfy the constraints. Additional SLOs follow the same convention (Appendix~\ref{app:surrogate-details}). Completed runs, including SLO violations, provide targets for predicting means $\mu_j(x)$ and standard deviations $\sigma_j(x)$. Execution probability $p_{\mathrm{exec}}(x)$ uses smoothed, similarity-weighted success and failure counts. Failed runs update this estimate but supply no energy or latency targets.

\subsection{Acquisition Function}
\label{sec:eco-acquisition}

The acquisition function weighs expected energy improvement by feasibility and evaluation cost. Under an independence approximation for the constraint predictions, SLO feasibility is estimated as $p_{\mathrm{SLO}}(x)=\prod_{j\in\mathcal J}\Phi(-\mu_j(x)/\sigma_j(x))$, where $\mathcal J$ contains all applicable service constraints and $\Phi$ is the standard normal cumulative distribution function. This probability guides exploration; measured outcomes determine final feasibility.

\ctwobo{} scores feasibility-weighted expected energy improvement per GPU-hour as
\begin{equation}
 a(x)=\frac{\mathbb E[(E^{\star}-E(x))_+]\,p_{\mathrm{SLO}}(x)\,p_{\mathrm{exec}}(x)}{\widehat c(x)}.
 \label{eq:eco-acquisition}
\end{equation}
Here $E^{\star}$ is the best measured feasible energy, $(u)_+=\max(u,0)$, and $\widehat c(x)$ is predicted trial cost in GPU-hours. The expectation uses the log-normal energy posterior. Until a feasible configuration is found, \ctwobo{} omits the improvement term and scores feasibility per unit cost. The initial cost estimate accounts for GPU allocation and trial lifecycle costs, including configuration switching. A GP corrects it using observed GPU-hours from successful and failed trials. Candidates whose predicted cost exceeds the remaining budget are excluded from the next trial.

% Declared in the method section to place the table at the top of page 6.
% Operating-point evidence: MOE_DVFS experiment-records/2026-09-16-a6000-progress/PLAN.json
% (also inherited by tools/prepare_a6000_8gpu.py), and the A100
% results/dbo_on_1256_v3/source/bo_dse/official.py frequency/power defaults.
% DBO threshold evidence: the A6000 PLAN and A100 dbo_on_1256_v3/PLAN.json.
\begin{table}[!t]
\vspace{-10pt}
\caption{Configuration ranges on A100 and A6000.}
\label{tab:design-space}
\centering
\footnotesize
\setlength{\tabcolsep}{4pt}
\begin{tabularx}{.95\linewidth}{@{}l >{\centering\arraybackslash}X >{\centering\arraybackslash}X@{}}
\toprule
Parameter & A100-SXM4 & A6000 \\
\midrule
Allocation $(N_A,N_F)$ & \multicolumn{2}{c@{}}{$1\leq N_F\leq N_A$, $N_A+N_F\leq G$, $N_F\mid N_A$} \\
Parallelism $(\mathrm{TP}_r,\mathrm{DP}_r,\mathrm{EP}_F)$ & \multicolumn{2}{c@{}}{$\mathrm{TP}_r\mid N_r$, $\mathrm{DP}_r=N_r/\mathrm{TP}_r$, $\mathrm{EP}_F=N_F$} \\
Decode/prefill thresholds & \multicolumn{2}{c@{}}{$(16,256)$, $(32,512)$, $(64,1024)$} \\
\midrule
Frequency / power cap & \shortstack{1050, 1170, 1290, 1350, 1410 MHz\\200, 250, 300, 350, 400 W/GPU} & \shortstack{1560, 1740, 1920, 2010, 2100 MHz\\200, 250, 300 W/GPU} \\
\bottomrule
\end{tabularx}
\vspace{-14pt}
\end{table}

\subsection{Candidate Filtering and Selection}
\label{sec:eco-selection}
\label{sec:eco-checks}
\label{sec:eco-feedback}

\textbf{Filter candidates and schedule measurements.}
\ctwobo{} removes configurations that violate static constraints; for example, 3A+2F violates $N_F\mid N_A$. Execution success must still be established by measurement. After the reference, \ctwobo{} interleaves acquisition choices with untried structures and single-control probes, such as varying only the Attention clock to estimate its stage-time response. Power-cap probes require evidence that the cap could affect execution. Appendix~\ref{app:search-policy} specifies the measurement policy.

\textbf{Rank candidates and select the final configuration.}
\ctwobo{} defers candidates whose optimistic predictions violate an SLO or exceed the best measured feasible energy by a margin, then measures the remaining candidate with the highest acquisition score (Equation~\ref{eq:eco-acquisition}). Periodic tests, model updates, and an empty active pool trigger reconsideration, preventing permanent exclusion from early prediction errors. ECO updates its predictive models using available trial measurements and execution outcomes. It returns the lowest-energy measured feasible configuration, or reports none, and freezes the settings before the independent held-out replay. Appendix~\ref{app:eligibility} specifies measurement eligibility and aggregation; held-out measurements do not update search.
 
% A6000 comparison with one panel per model and a shared caption.
\begin{figure}[!t]
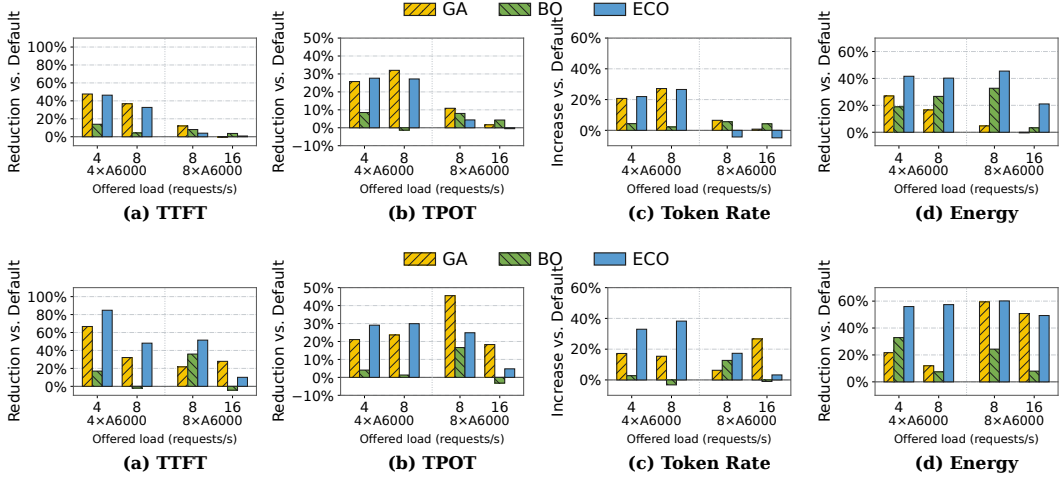

  \vspace{-6pt}
  \centering
  \input{figures/a6000_qwen_comparison_figure}
  \par\vspace{-6pt}
  \input{figures/a6000_deepseek_comparison_figure}
  \setlength{\abovecaptionskip}{4pt}
  \caption{A6000 vs. Default on the same held-out requests: Qwen (top), DeepSeek (bottom). Higher is better.}
  \label{fig:a6000-comparison}
  \vspace{-10pt}
\end{figure}

\section{Experimental Evaluation}
\label{sec:experiments}
% Present the setup directly under the evaluation heading to avoid two
% consecutive headings before the first experimental paragraph.
\label{sec:experiment-setup}

\begin{figure}[t]
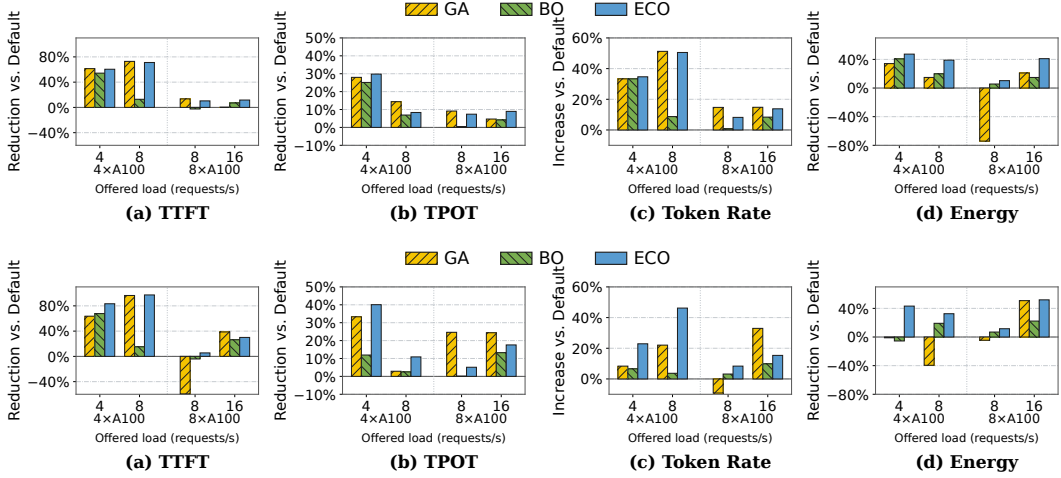

  \vspace{-6pt}
  \centering
  \input{figures/a100_qwen_comparison_figure}
  \par\vspace{-6pt}
  \input{figures/a100_deepseek_comparison_figure}
  \setlength{\abovecaptionskip}{4pt}
  \caption{A100 vs. Default on the same held-out requests: Qwen (top), DeepSeek (bottom). Higher is better.}
  \label{fig:a100-comparison}
  \vspace{-15pt}
\end{figure}

We evaluate Qwen3.6-35B-A3B~\citep{qwen36_35b_a3b} and DeepSeek-V2-Lite-Chat~\citep{deepseekv2} using vLLM~0.26.0~\citep{kwon2023pagedattention} with AFD on A100-SXM4 (80\,GB) and A6000 (48\,GB) GPUs.
On each platform, both models use four-GPU pools at 4/8 requests/s and eight-GPU pools at 8/16 requests/s.
ECO, constrained BO, and GA independently search the same fixed 200-request workload constructed from the Splitwise Azure trace~\citep{patel2024splitwise}.
For each scenario, we repeat each method's search with multiple random seeds and report averages across these independent runs.
Each run allows 16 physical search trials, including initialization probes and failures, plus one shared preparation evaluation charged separately (17 evaluations per run in total).
Search and held-out replays use real text prompts from replay files; request identities, prompts, token lengths, and arrival schedules are identical across methods within each split.
BO uses a constant energy prior.
Default uses the full GPU pool with equal Attention and FFN groups, while selected configurations may use fewer GPUs.
SLO thresholds are fixed before search at 105\% of historical Default latency and 95\% of historical Default throughput, serving as latency ceilings and throughput floors, respectively.
TTBT constraints apply where specified, and all thresholds remain fixed during held-out evaluation.
In each run, each method freezes its lowest-energy measured feasible configuration before replaying 200 disjoint held-out requests.
Default is rerun on the same held-out requests and arrival schedule to provide the comparison denominators.
Figures~\ref{fig:a6000-comparison} and~\ref{fig:a100-comparison} report p90 TTFT/TPOT, output tokens/s, and serving energy from these replays; held-out results do not affect configuration selection.
Appendix~\ref{app:protocol} details the measurement protocol, and Appendix~\ref{app:baseline-settings} specifies BO and GA settings.

% Layout anchor: the A100 result pair follows on page 8.
% A100 comparison with one panel per model and a shared caption.

% Queue the search figure after both platform comparison pairs.
% Keep the search figure above a single ablation figure with three panels.
\newsavebox{\searchtrialplot}
\newsavebox{\energyablationplot}
\begin{figure}[t]
  \vspace{-12pt}
  \centering
  \sbox{\searchtrialplot}{\includegraphics[width=\linewidth]{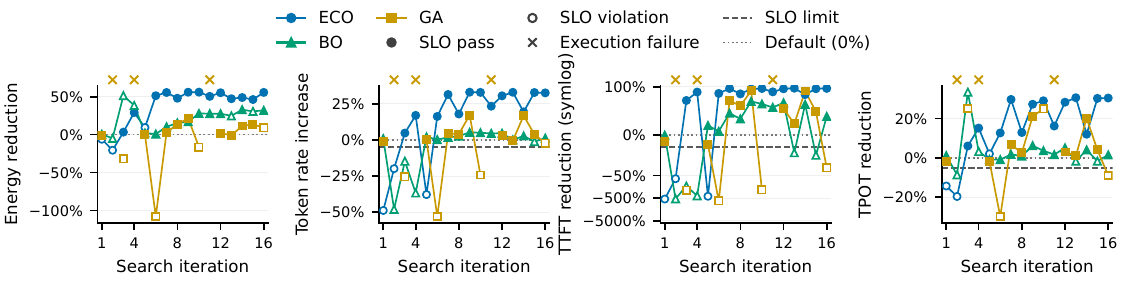}}
  \usebox{\searchtrialplot}
  \vspace{-16pt}
  \caption{Search trajectories for DeepSeek on four A6000 GPUs at 4 requests/s. Trials use the same search workload; changes are relative to Default (higher is better).}
  \label{fig:search-trials}
  \par
  \sbox{\energyablationplot}{\includegraphics[width=.414\linewidth]{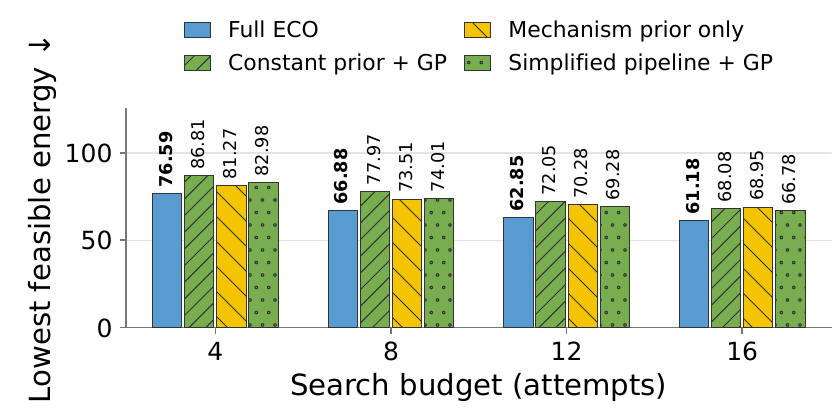}}
  \begin{minipage}[t]{.414\linewidth}
    \vspace{0pt}
    \input{figures/ablation_energy_figure}
  \end{minipage}\hfill
  \begin{minipage}[t]{.281\linewidth}
    \vspace{0pt}
    \input{figures/joint_tuning_ablation_figure}
  \end{minipage}\hfill
  \begin{minipage}[t]{.281\linewidth}
    \vspace{0pt}
    \input{figures/cost_aware_ablation_figure}
  \end{minipage}
  \vspace{-10pt}
  \caption{Ablations on DeepSeek with four A100 GPUs. Left to right: energy model, joint versus sequential tuning, and cost-aware acquisition.}
  \label{fig:ablations}
  \vspace{-15pt}
\end{figure}

\subsection{Results}
\label{sec:results}
\label{sec:energy-results}
\label{sec:results-a6000}
\label{sec:results-a100}
\label{sec:search-behavior}

% Paragraph role: overall energy results, platform differences, and service trade-offs.
\textbf{Energy savings across platforms.}
ECO reduces energy relative to Default in all 16 scenarios and achieves the lowest energy among the three search methods in 15 (Figures~\ref{fig:a6000-comparison} and~\ref{fig:a100-comparison}). Across all 16 scenarios on A6000 and A100, its mean energy reduction is 40.5\% and its mean output-token-rate increase is 20.7\%, with each scenario weighted equally. Across the eight A6000 scenarios, selected feasible energy averages 33.1\% below BO and 25.8\% below GA. On A6000, Qwen's four-GPU selections improve latency and token rate, whereas its eight-GPU selections operate near the throughput floor; DeepSeek improves the plotted service metrics, but GA slightly outperforms ECO's energy at the higher eight-GPU load. On A100, ECO achieves the lowest energy in every setting while improving latency and token rate over Default. GA sometimes improves service metrics further without matching ECO's energy; for eight-GPU Qwen at the lower load, its higher token rate comes with greater energy use than Default.

% Paragraph role: search efficiency measured against a common feasible-energy target.
\textbf{Fewer attempts to reach the energy target.}
Figure~\ref{fig:search-trials} shows 16 search attempts per method for DeepSeek on four A6000 GPUs at 4 requests/s. Using GA's best feasible energy, 30.93\,kJ, as a common target, ECO reaches a feasible configuration at or below that energy after 4 attempts, versus 10 for BO and 9 for GA. Their lowest feasible energies are 17.39, 26.50, and 30.93\,kJ, respectively. Appendix~\ref{app:supplementary-search} details completed, SLO-feasible, and failed attempts. These trajectories compare search outcomes; matched ablations are needed to attribute differences to individual components.

\subsection{Ablation Study}
\label{sec:ablation}

% Paragraph role: observed energy-model differences, then mechanism interpretation.
\textbf{The full surrogate finds lower-energy configurations.}
All three ablations use DeepSeek on four A100 GPUs. Full ECO finds the lowest feasible energy at every budget in Figure~\ref{fig:ablations} (left). At budget 16, Full ECO uses 10.1\% less energy than Constant prior + GP, which overtakes Mechanism prior only; Simplified pipeline + GP remains 9.2\% above Full ECO. These variants respectively replace the mechanism prior with a constant, remove residual correction, or replace FIFO scheduling with a bottleneck approximation. The trends suggest that the prior guides early search, residual learning corrects persistent error, and pipeline scheduling improves energy estimates.

% Paragraph role: load-dependent joint-tuning gains, then search-space explanation.
\textbf{Joint tuning yields additional energy savings.}
Joint tuning saves 40.2\% and 18.1\% relative to each load's Default at 4 and 8 requests/s in Figure~\ref{fig:ablations} (middle), compared with 30.7\% and 14.6\% for sequential tuning. The additional energy reduction over sequential tuning narrows from 13.7\% to 4.1\% at higher load. Sequential tuning fixes the deployment structure before adjusting operating controls, restricting the combinations considered in the second step; joint tuning retains the alternatives.

% Paragraph role: budget-dependent energy differences, then cost-normalization rationale.
\textbf{Cost awareness lowers energy under GPU-hour budgets.}
At 4 requests/s, Full ECO uses 21.1\% and 10.8\% less energy at budgets of 2.0 and 2.5 GPU-hours in Figure~\ref{fig:ablations} (right), despite matching No-cost after six attempts. No-cost removes acquisition cost normalization.

\subsection{Key Findings}
\label{sec:configuration-findings}

% Paragraph role: connect reuse at the tested load points to reconfiguration decisions.
\textbf{Finding 1. Load growth can preserve the efficient allocation and clocks.}
On four A6000 GPUs, both models retain their selected GPU counts and clocks when the tested request rate doubles (Table~\ref{tab:appendix-configs}). This suggests checking whether the existing allocation and clocks remain feasible before incurring a structural reconfiguration. Operating controls such as power caps can still require adjustment within that structure.

% Paragraph role: derive independent stage scaling from the observed asymmetric selections.
\textbf{Finding 2. Efficient AFD allocations can be strongly asymmetric.}
Under the eight-GPU budget, both models select Attention-heavy deployments. At higher load, DeepSeek adds Attention capacity while retaining its FFN pool. These selections motivate scaling the two GPU groups independently and comparing allocation ratios jointly with operating controls; they do not isolate the benefit of any single configuration choice.

% Paragraph role: turn narrow measured feasibility margins into a search-objective implication.
\textbf{Finding 3. Minimum measured energy can leave little service headroom.}
Both eight-GPU Qwen selections have less than 1\% throughput headroom above the SLO floor. A deployment requiring additional service margin should encode that margin in the search constraints. Doing so changes the feasible set and can change the selected configuration and its energy savings; feasibility on the fixed search workload alone does not establish robustness to workload shifts or higher demand.

\vspace{-6pt}
\section{Related Work}
\label{sec:related-work}
\vspace{-6pt}

% Paragraph role: energy literature; acknowledge the closest joint AFD/DVFS system.
\textbf{Power and energy management.}
POLCA supports cluster power oversubscription~\citep{patel2024polca}; DynamoLLM jointly adjusts instance count, parallelism, and frequency~\citep{stojkovic2024dynamollm}; DualScale coordinates placement and frequency control in prefill/decode disaggregation~\citep{basit2026dualscale}. AFlex is closest to our setting, combining Attention--FFN provisioning and frequency control with offline operator profiles, global scheduling, and adaptive pipelining and batching~\citep{hu2026aflex}. Runtime approaches include throttLL'eM's iteration-level performance prediction~\citep{kakolyris2024throttllem} and GreenLLM's separate prefill optimization and decode feedback control~\citep{liu2025greenllm}. ECO uses AFD hierarchy and pipeline dependencies to select measurements under a limited search budget, returning a measured static configuration for the target workload.

% Paragraph role: search literature; position the combination without claiming BO primitives as new.
\textbf{Automatic configuration search.}
CherryPick uses BO for cloud configuration selection~\citep{alipourfard2017cherrypick}. For LLM inference, SCOOT combines single- and multi-objective BO, known-constraint pruning, hidden-constraint modeling, and parallel suggestions~\citep{cheng2025scoot}. ECO builds on constrained BO for unknown feasibility~\citep{gelbart2014constraints,gardner2014constraints} and cost-aware BO for varying evaluation duration~\citep{snoek2012practical}. Its AFD-specific energy prior conditions stage responses on deployment structure and propagates pipeline timing before residual correction. Our characterization and model ablation evaluate the information this prior supplies.

\vspace{-6pt}
\section{Conclusion}
\label{sec:conclusion}
\vspace{-6pt}

ECO combines a calibrated stage-and-pipeline energy prior, residual learning, and cost-aware constrained Bayesian optimization to jointly select AFD deployment structures and operating controls under a limited measurement budget. Across 16 A6000 and A100 scenarios, frozen configurations evaluated on disjoint requests reduce serving energy by 40.5\% and increase output token rate by 20.7\% on average relative to Default while meeting SLOs. Future work should evaluate separate adaptation schedules end-to-end and quantify their mechanisms through controlled attribution and stage-waiting measurements.

\clearpage
\section*{AI Use Statement}

We used generative AI tools to assist with implementing portions of the code and polishing the writing of this manuscript. The authors take responsibility for the final code, experimental results, and manuscript, including content produced with AI assistance.

\section*{Ethics Statement}

This work studies energy-efficient LLM serving while maintaining service-level objectives. Our experiments use real text prompts stored in the replay trace files, with request timing and token-length metadata from the Splitwise Azure trace. The study focuses on serving efficiency, with energy consumption and service quality evaluated together.

\section*{Reproducibility Statement}

Section~\ref{sec:experiment-setup} describes the experimental setup, Appendix~\ref{app:protocol} specifies the workload and measurement protocol, and Appendix~\ref{app:search} details the search procedures and baseline settings. These sections document the implementation and configuration used in the experiments under the reported hardware and software settings.

\clearpage
\bibliography{references}

@inproceedings{patel2024polca,
  title = {Characterizing Power Management Opportunities for {LLMs} in the Cloud},
  author = {Patel, Pratyush and Choukse, Esha and Zhang, Chaojie and Goiri, {\'I}{\~n}igo and Warrier, Brijesh and Mahalingam, Nithish and Bianchini, Ricardo},
  booktitle = {Proceedings of the ACM International Conference on Architectural Support for Programming Languages and Operating Systems (ASPLOS)},
  year = {2024},
  url = {https://www.microsoft.com/en-us/research/publication/characterizing-power-management-opportunities-for-llms-in-the-cloud/}
}

@misc{stojkovic2024dynamollm,
  title = {{DynamoLLM}: Designing {LLM} Inference Clusters for Performance and Energy Efficiency},
  author = {Stojkovic, Jovan and Zhang, Chaojie and Goiri, {\'I}{\~n}igo and Torrellas, Josep and Choukse, Esha},
  year = {2024},
  eprint = {2408.00741},
  archivePrefix = {arXiv},
  url = {https://arxiv.org/abs/2408.00741}
}

@misc{wu2026disaggregation,
  title = {How Far Can Disaggregation Go? A Design-Space Exploration of Attention-{FFN} Disaggregation for Efficient {MoE LLM} Serving},
  author = {Wu, Hanjiang and Bambhaniya, Abhimanyu Rajeshkumar and Banerjee, Sarbartha and Khare, Tuhin and Srinivasan, Sudarshan and Subramanian, Suvinay and Kundu, Souvik and Kumar, Madhu and Elavazhagan, Midhilesh and Won, William and Yazdanbakhsh, Amir and Krishna, Tushar},
  year = {2026},
  eprint = {2605.28302},
  archivePrefix = {arXiv},
  url = {https://arxiv.org/abs/2605.28302}
}

@misc{basit2026dualscale,
  title = {{DualScale}: Energy-Efficient Disaggregated {LLM} Serving via Phase-Aware Placement and {DVFS}},
  author = {Basit, Omar and Liu, Yunzhao and Kong, Z. Jonny and Hu, Y. Charlie},
  year = {2026},
  eprint = {2602.18755},
  archivePrefix = {arXiv},
  url = {https://arxiv.org/abs/2602.18755v3}
}

@misc{hu2026aflex,
  title = {Energy-Efficient {LLM} Serving via Disaggregated Attention--{FFN} and Flexible Frequency Scaling},
  author = {Hu, Cunchen and Xu, Liangliang and Liu, Tian and Lyu, Min and Li, Yongkun and Wang, Sa and Quan, Shuo and Yang, Yanan and Tang, Wenda and Wang, Yiduo and Yu, Fu and Wu, Jie},
  year = {2026},
  eprint = {2608.01891},
  archivePrefix = {arXiv},
  url = {https://arxiv.org/abs/2608.01891}
}

@inproceedings{alipourfard2017cherrypick,
  title = {{CherryPick}: Adaptively Unearthing the Best Cloud Configurations for Big Data Analytics},
  author = {Alipourfard, Omid and Liu, Hongqiang Harry and Chen, Jianshu and Venkataraman, Shivaram and Yu, Minlan and Zhang, Ming},
  booktitle = {14th USENIX Symposium on Networked Systems Design and Implementation (NSDI 17)},
  year = {2017},
  pages = {469--482},
  publisher = {USENIX Association},
  url = {https://www.usenix.org/conference/nsdi17/technical-sessions/presentation/alipourfard}
}

@misc{gelbart2014constraints,
  title = {Bayesian Optimization with Unknown Constraints},
  author = {Gelbart, Michael A. and Snoek, Jasper and Adams, Ryan P.},
  year = {2014},
  eprint = {1403.5607},
  archivePrefix = {arXiv},
  url = {https://arxiv.org/abs/1403.5607}
}

@inproceedings{patel2024splitwise,
  title = {{Splitwise}: Efficient Generative {LLM} Inference Using Phase Splitting},
  author = {Patel, Pratyush and Choukse, Esha and Zhang, Chaojie and Shah, Aashaka and Goiri, I{\~n}igo and Maleki, Saeed and Bianchini, Ricardo},
  booktitle = {Proceedings of the International Symposium on Computer Architecture (ISCA)},
  year = {2024},
  url = {https://www.microsoft.com/en-us/research/publication/splitwise-efficient-generative-llm-inference-using-phase-splitting/}
}

@misc{qwen36_35b_a3b,
  title = {{Qwen3.6-35B-A3B}: Agentic Coding Power, Now Open to All},
  author = {{Qwen Team}},
  year = {2026},
  month = apr,
  url = {https://qwen.ai/blog?id=qwen3.6-35b-a3b}
}

@misc{deepseekv2,
  title = {{DeepSeek-V2}: A Strong, Economical, and Efficient Mixture-of-Experts Language Model},
  author = {{DeepSeek-AI}},
  year = {2024},
  eprint = {2405.04434},
  archivePrefix = {arXiv},
  primaryClass = {cs.CL},
  url = {https://arxiv.org/abs/2405.04434}
}

@misc{zhu2025megascale,
  title = {{MegaScale-Infer}: Serving Mixture-of-Experts at Scale with Disaggregated Expert Parallelism},
  author = {Zhu, Ruidong and Jiang, Ziheng and Jin, Chao and Wu, Peng and Stuardo, Cesar A. and Wang, Dongyang and Zhang, Xinlei and Zhou, Huaping and Wei, Haoran and Cheng, Yang and Xiao, Jianzhe and Zhang, Xinyi and Liu, Lingjun and Lin, Haibin and Chang, Li-Wen and Ye, Jianxi and Yu, Xiao and Liu, Xuanzhe and Jin, Xin and Liu, Xin},
  year = {2025},
  eprint = {2504.02263},
  archivePrefix = {arXiv},
  url = {https://arxiv.org/abs/2504.02263}
}

@inproceedings{cheng2025scoot,
  title = {{SCOOT}: {SLO}-Oriented Performance Tuning for {LLM} Inference Engines},
  author = {Cheng, Ke and Wang, Zhi and Hu, Wen and Yang, Tiannuo and Li, Jianguo and Zhang, Sheng},
  booktitle = {Proceedings of the ACM on Web Conference 2025},
  year = {2025},
  pages = {829--839},
  url = {https://arxiv.org/abs/2408.04323}
}

@inproceedings{snoek2012practical,
  title = {Practical Bayesian Optimization of Machine Learning Algorithms},
  author = {Snoek, Jasper and Larochelle, Hugo and Adams, Ryan P.},
  booktitle = {Advances in Neural Information Processing Systems},
  volume = {25},
  year = {2012},
  url = {https://papers.nips.cc/paper/2012/hash/05311655a15b75fab86956663e1819cd-Abstract.html}
}

@inproceedings{kwon2023pagedattention,
  title = {Efficient Memory Management for Large Language Model Serving with {PagedAttention}},
  author = {Kwon, Woosuk and Li, Zhuohan and Zhuang, Siyuan and Sheng, Ying and Zheng, Lianmin and Yu, Cody Hao and Gonzalez, Joseph E. and Zhang, Hao and Stoica, Ion},
  booktitle = {Proceedings of the 29th Symposium on Operating Systems Principles (SOSP)},
  year = {2023},
  doi = {10.1145/3600006.3613165},
  url = {https://arxiv.org/abs/2309.06180}
}

@inproceedings{yu2022orca,
  title = {{Orca}: A Distributed Serving System for {Transformer-Based} Generative Models},
  author = {Yu, Gyeong-In and Jeong, Joo Seong and Kim, Geon-Woo and Kim, Soojeong and Chun, Byung-Gon},
  booktitle = {16th USENIX Symposium on Operating Systems Design and Implementation (OSDI 22)},
  year = {2022},
  pages = {521--538},
  publisher = {USENIX Association},
  url = {https://www.usenix.org/conference/osdi22/presentation/yu}
}

@inproceedings{agrawal2024sarathi,
  title = {Taming Throughput-Latency Tradeoff in {LLM} Inference with {Sarathi-Serve}},
  author = {Agrawal, Amey and Kedia, Nitin and Panwar, Ashish and Mohan, Jayashree and Kwatra, Nipun and Gulavani, Bhargav and Tumanov, Alexey and Ramjee, Ramachandran},
  booktitle = {18th USENIX Symposium on Operating Systems Design and Implementation (OSDI 24)},
  year = {2024},
  pages = {117--134},
  publisher = {USENIX Association},
  url = {https://www.usenix.org/conference/osdi24/presentation/agrawal}
}

@inproceedings{rajbhandari2022deepspeedmoe,
  title = {{DeepSpeed-MoE}: Advancing Mixture-of-Experts Inference and Training to Power Next-Generation {AI} Scale},
  author = {Rajbhandari, Samyam and Li, Conglong and Yao, Zhewei and Zhang, Minjia and Aminabadi, Reza Yazdani and Awan, Ammar Ahmad and Rasley, Jeff and He, Yuxiong},
  booktitle = {Proceedings of the 39th International Conference on Machine Learning},
  volume = {162},
  pages = {18332--18346},
  year = {2022},
  publisher = {PMLR},
  url = {https://proceedings.mlr.press/v162/rajbhandari22a.html}
}

@inproceedings{gardner2014constraints,
  title = {Bayesian Optimization with Inequality Constraints},
  author = {Gardner, Jacob R. and Kusner, Matt J. and Xu, Zhixiang and Weinberger, Kilian Q. and Cunningham, John P.},
  booktitle = {Proceedings of the 31st International Conference on Machine Learning},
  volume = {32},
  pages = {937--945},
  year = {2014},
  publisher = {PMLR},
  url = {https://proceedings.mlr.press/v32/gardner14.html}
}

@book{rasmussen2006gpml,
  title = {Gaussian Processes for Machine Learning},
  author = {Rasmussen, Carl Edward and Williams, Christopher K. I.},
  publisher = {MIT Press},
  year = {2006},
  isbn = {978-0-262-18253-9},
  url = {https://gaussianprocess.org/gpml/}
}

@misc{kakolyris2024throttllem,
  title = {{SLO}-aware {GPU} Frequency Scaling for Energy Efficient {LLM} Inference Serving},
  author = {Kakolyris, Andreas Kosmas and Masouros, Dimosthenis and Vavaroutsos, Petros and Xydis, Sotirios and Soudris, Dimitrios},
  year = {2024},
  eprint = {2408.05235},
  archivePrefix = {arXiv},
  url = {https://arxiv.org/abs/2408.05235}
}

@misc{liu2025greenllm,
  title = {{GreenLLM}: {SLO}-Aware Dynamic Frequency Scaling for Energy-Efficient {LLM} Serving},
  author = {Liu, Qunyou and Huang, Darong and Zapater, Marina and Atienza, David},
  year = {2025},
  eprint = {2508.16449},
  archivePrefix = {arXiv},
  url = {https://arxiv.org/abs/2508.16449}
}
\bibliographystyle{iclr2027_conference}
\clearpage
\appendix
\section{Appendix}
\label{app:protocol}

% Paragraph role: opening; anchor the supplement to the current static method.
The static configurations in Sections~\ref{sec:problem-formulation}--\ref{sec:experiments} fix the deployment structure and operating controls for each serving replay. We use $A$ for attention and $F$ for FFN computation, with E denoting expert GPUs in deployment descriptions.

\subsection{Workloads and Deployment Settings}
\label{app:workloads}

% Paragraph role: design; make the scenario matrix explicit.
We evaluate 16 scenarios: two models on two GPU platforms, each with the four GPU-budget/load pairs in Table~\ref{tab:appendix-scenarios} (Section~\ref{sec:experiment-setup}). The GPU budget $G$ bounds the selected allocation, $N_A+N_F\leq G$. A configuration using six GPUs can therefore be selected under the eight-GPU budget, $G=8$.

\begin{table}[htbp]
\caption{Scenario grid, repeated for both models on both platforms. Each scenario uses a fixed 200-request search workload constructed from the Splitwise Azure trace, eight warmup requests, and a separate 200-request held-out replay. TTFT, TPOT, and output token rate are constrained in every scenario.}
\label{tab:appendix-scenarios}
\centering
\small
\begin{tabular}{@{}rrrrc@{}}
\toprule
GPU budget $G$ & Load (requests/s) & Default $N_A$ & Default $N_F$ & TTBT constraint \\
\midrule
4 & 4 & 2 & 2 & No \\
4 & 8 & 2 & 2 & Yes \\
8 & 8 & 4 & 4 & Yes \\
8 & 16 & 4 & 4 & Yes \\
\bottomrule
\end{tabular}
\end{table}

% Paragraph role: design; specify what the trace replay preserves.
We construct the replay workload from request arrival times and input/output token lengths recorded in the Splitwise Azure trace~\citep{patel2024splitwise}. Both the search and held-out workloads use real text prompts read from the replay trace files. The requested output length is capped at 128 tokens, using the smaller of the recorded output length and this cap. Generation uses temperature zero and disables end-of-sequence termination so that early stopping does not change the requested workload. Requests follow a predetermined submission schedule, allowing arrivals to overlap with ongoing generation. All methods in a scenario receive the same prompt texts and request schedule within each split; the held-out replay uses separate request identities with the matched arrival schedule specified in Section~\ref{sec:experiment-setup}.

% Paragraph role: design; separate search measurements from the independent main evaluation.
For each scenario, we repeat ECO, constrained BO, and GA searches with multiple random seeds, maintaining independent search histories for each run. The fixed search workload defines the objective for physical configuration trials. ECO also uses trial measurements for stage profiling, model updates, and later search decisions. The eight warmup requests come from the search workload and are excluded from serving metrics. SLO thresholds are fixed from the historical Default before search. In each run, we freeze the selected configuration before the independent held-out replay; the reported comparisons average results across these independent search seeds. Held-out observations never update the optimizer or change its selection. Matched arrival schedules isolate the change in request sample; this replay does not test shifts in offered load. Replay files must retain source request identities: changing a file name or arrival rate does not create a disjoint request set.

% Paragraph role: design; clarify how the published configuration space is interpreted.
Table~\ref{tab:design-space} specifies the operating-control grids. Operating controls apply per GPU within each role, so attention and expert GPUs may use different settings. The derived parallel degrees satisfy $\mathrm{DP}_r=N_r/\mathrm{TP}_r$ and $\mathrm{EP}_F=N_F$. These algebraic conditions are checked together with model divisibility and backend support. DBO uses two microbatches when its decode or prefill threshold is reached; enabling DBO does not imply that every batch is split. Default enables DBO with thresholds $(32,512)$ and uses 1410\,MHz/400\,W on A100 or 2100\,MHz/300\,W on A6000.

\subsection{Latency, Throughput, and SLOs}
\label{app:metrics}

% Paragraph role: definition; make client timing and percentile aggregation reproducible.
Latency is measured at the serving client. For request $i$, let $a_i$ be its actual submission time, $u_i$ its first-token arrival time, $d_i$ its response completion time, and $n_i$ its completed output-token count. The request-level metrics, in seconds, are
\begin{equation}
 \mathrm{TTFT}_i=u_i-a_i,\qquad
 \mathrm{TPOT}_i=\frac{d_i-u_i}{\max(n_i-1,1)}.
 \label{eq:appendix-request-latency}
\end{equation}
We compute $L_{\mathrm{TTFT}}$ and $L_{\mathrm{TPOT}}$ as the p90 values across requests, converting to milliseconds where reported. Quantiles use linear interpolation between adjacent sorted observations: for $m$ values and quantile $q$, the fractional zero-based index is $q(m-1)$. TPOT is computed for each request before taking the percentile.

% Paragraph role: definition; distinguish the optional token-interval measurement.
TTBT measures individual token-arrival intervals within a request. Let $u_{i,j}$ be the client arrival time of output token $j$. We pool the intervals $u_{i,j+1}-u_{i,j}$ across requests and take their p90. Tokens delivered in the same streaming event share a timestamp, so stream coalescing affects this measurement. TPOT instead averages each request's post-first-token duration. Scenarios with a TTBT constraint check both metrics separately.

% Paragraph role: definition; tie output throughput and thresholds to the main objective.
The serving interval is $T=\max_i d_i-\min_i a_i$, and output token rate is $R=\sum_i n_i/T$. The superscript $b$ denotes the baseline used only to set SLO thresholds, measured in the historical Default run for the same model, GPU platform, GPU budget, and offered load. We set the fixed SLO thresholds from these baseline latencies and output token rate:
\begin{equation}
 \begin{alignedat}{2}
 \tau_{\mathrm{TTFT}} &= 1.05\,L_{\mathrm{TTFT}}^{b},
 &\qquad \tau_{\mathrm{TPOT}} &= 1.05\,L_{\mathrm{TPOT}}^{b},\\[4pt]
 \tau_{\mathrm{TTBT}} &= 1.05\,L_{\mathrm{TTBT}}^{b},
 &\qquad R_{\min} &= 0.95\,R^{b}.
 \end{alignedat}
 \label{eq:appendix-slos}
\end{equation}
The TTBT threshold adds an acceptance condition for the rows marked Yes in Table~\ref{tab:appendix-scenarios}. All applicable thresholds must be met and remain fixed throughout search and held-out replay.

\subsection{Energy Accounting and Result Normalization}
\label{app:energy}

% Paragraph role: definition; clarify the energy boundary and what fewer GPUs means.
Serving energy integrates power measured through the NVIDIA Management Library (NVML) from the first request submission to the last response completion:
\begin{equation}
 E(x;\mathcal W)=\int_0^{T(x;\mathcal W)}\sum_{g\in\mathcal G(x)}P_g(t)\,dt,
 \label{eq:dse-energy}
\end{equation}
where $\mathcal G(x)$ is the active GPU set, $P_g(t)$ is measured GPU power, and $T$ is the same interval used for output token rate. Energy includes idle and waiting periods on the selected GPUs and excludes GPUs outside that allocation. Model loading and warmup precede the serving window. Power caps limit the operating point; the energy calculation uses the measured power samples.

% Paragraph role: design; explain measurement validity and search-cost accounting.
A usable trial must complete all requests and pass output and telemetry checks. Request logs determine the measured interval and output count; power records establish coverage of that interval. Failed launches, incomplete responses, and invalid measurements consume search attempts but cannot qualify as low-energy results. Search cost includes all reserved GPU time over the full trial lifecycle, including failed trials and any overrun by an ongoing trial.

% Paragraph role: definition; reproduce the main figures' directions and denominators.
We rerun Default on the same 200 held-out requests and arrival schedule as each frozen search-selected configuration. Let $M^{D,h}$ denote a metric from this Default replay, and $M^{x,h}$ the corresponding metric for the selected configuration. For each seed, the energy or latency percentage change is $100(1-M^{x,h}/M^{D,h})$; for output token rate it is $100(R^{x,h}/R^{D,h}-1)$. Positive values indicate improvement. The historical measurements $M^b$ in Section~\ref{app:metrics} set the SLO thresholds before search; the held-out Default measurements supply the comparison denominators. Consequently, the fixed latency threshold appears at $100(1-\tau/M^{D,h})$ and the throughput floor at $100(R_{\min}/R^{D,h}-1)$, which need not equal $-5\%$. Within each search run, all metrics for a method/scenario come from the held-out replay of that run's frozen configuration.

For each scenario $q$, we average percentage changes over the independent search seeds, denoting the mean energy reduction by $\overline{\Delta E}_q$ and the mean output-token-rate increase by $\overline{\Delta R}_q$. We then summarize all 16 model/platform/GPU-budget/load scenarios with equal weight: $\frac{1}{16}\sum_{q=1}^{16}\overline{\Delta E}_q\approx40.5\%$ and $\frac{1}{16}\sum_{q=1}^{16}\overline{\Delta R}_q\approx20.7\%$. Each four- and eight-GPU scenario contributes once to these averages.

\section{Search and Selection Details}
\label{app:search}

\subsection{Measurement Outcomes and Configuration Eligibility}
\label{app:eligibility}

% Paragraph role: design; spell out the three outcomes used by the models.
For each method and random seed, the 16-attempt search allowance counts all in-search candidate evaluations, including initialization probes and failures. Preparation is charged separately as one evaluation and its GPU-hour cost; a completed 16-attempt search therefore has 17 charged evaluations in total, using the accounting in Section~\ref{sec:problem-formulation}. Methods share the workload and service thresholds while maintaining independent search histories. Generic constrained BO uses the constant energy prior specified in Section~\ref{sec:experiment-setup}. ECO uses the stage-based prior and search rules in Section~\ref{sec:eco-methods}. Table~\ref{tab:appendix-outcomes} lists the model feedback and selection consequences for each outcome.

\begin{table}[htbp]
\caption{Use of trial outcomes in ECO. Feasibility is determined by the complete scenario-specific SLO rule.}
\label{tab:appendix-outcomes}
\centering
\small
\begin{tabularx}{\linewidth}{@{}>{\raggedright\arraybackslash}p{.27\linewidth}>{\raggedright\arraybackslash}X>{\raggedright\arraybackslash}X@{}}
\toprule
Outcome & Model feedback & Final-selection consequence \\
\midrule
Execution or measurement failure & Execution outcome and incurred cost; no energy or service target for the GPs. & Ineligible; still consumes an attempt. \\
Completed, SLO-violating trial & Measured energy, service metrics, execution success, and cost. & Retained in search history; SLO eligibility is checked at configuration level. \\
Completed, SLO-feasible trial & Measured energy, service metrics, execution success, and cost. & Can become the incumbent after configuration-level eligibility checks. \\
\bottomrule
\end{tabularx}
\end{table}

% Paragraph role: definition; preserve the main text's repeated-candidate rule.
Within each search run, ECO averages energy and service metrics across repeated attempts of the same configuration. Latency aggregation uses the per-run p90 values, rather than a percentile of pooled request samples. Eligibility requires every attempt to succeed and the averaged service metrics to meet all applicable SLOs. ECO returns the eligible configuration with the lowest mean energy; a low-energy trial that violates a constraint cannot by itself replace the feasible incumbent.

\subsection{Surrogate Implementation Details}
\label{app:surrogate-details}

Following Section~\ref{sec:eco-surrogates}, we fit the stage-time model separately for each measured deployment structure $s$. Within a fixed structure, the base model for stage $k$ at a measured operating point is
\begin{equation}
 \widehat s_k^0=b_k+a_k^p n_p+a_k^d n_d,
 \qquad b_k,a_k^p,a_k^d\geq0,
 \label{eq:eco-stage-time}
\end{equation}
where $n_p,n_d$ are prefill and decode tokens per microbatch. The coefficients are specific to the fixed structure and are fitted to its calibration stage records by nonnegative least squares. For control $u_j$ and reference $u_j^0$, the local response is $\widehat s_k(u_j)=s_k(u_j^0)(u_j/u_j^0)^{-\eta_{kj}}$. We fit $\eta_{kj}$ from paired log time ratios, keeping the deployment structure and all other operating controls fixed, as in Section~\ref{sec:eco-surrogates}. Multiple control changes combine these local responses unless calibration identifies a joint effect. Unidentified effects and extrapolation increase prior uncertainty.

When query-context records are available, the Attention stage model uses the total context length across queries to predict the slowest participating rank's service time. Routing imbalance adjusts FFN time, and transferred bytes scale dispatch and combine times. Positive calibration residuals provide an empirical margin against underprediction. In the pipeline simulator, calibration service weights distribute fitted stage totals across layers; missing layer records use uniform weights.

The residual GP takes normalized log operating controls, a one-hot encoding of deployment structure, and stage times normalized by the longest stage as inputs. It uses a Mat\'ern-$5/2$ kernel~\citep{rasmussen2006gpml}.

Service GPs predict normalized constraints:
\begin{equation}
 g_1=\log\frac{L_{\mathrm{TTFT}}}{\tau_{\mathrm{TTFT}}},\qquad
 g_2=\log\frac{L_{\mathrm{TPOT}}}{\tau_{\mathrm{TPOT}}},\qquad
 g_3=\log\frac{R_{\min}}{R}.
 \label{eq:eco-constraints}
\end{equation}
Feasibility requires each value to be nonpositive. Additional scenario-specific constraints, including TTBT, use the same normalization. Training includes completed SLO-violating runs, but excludes execution failures.

\subsection{Expected Improvement in Energy Units}
\label{app:ei}

% Paragraph role: derivation; expand the acquisition expectation without adding a model.
The energy GP predicts log energy, whereas the improvement term in Equation~\ref{eq:eco-acquisition} is measured in joules. Let $Z=\log E\sim\mathcal N(\mu_E,\sigma_E^2)$ at a candidate and let $E^\star>0$ be the current feasible incumbent energy. For $\sigma_E>0$, define $z=(\log E^\star-\mu_E)/\sigma_E$. Integrating the log-normal posterior gives
\begin{equation}
 \begin{aligned}
 \operatorname{EI}_E
 &=\int_{-\infty}^{\log E^\star}(E^\star-e^t)
       \frac{1}{\sigma_E}\phi\!\left(\frac{t-\mu_E}{\sigma_E}\right)\,dt\\
 &=E^\star\Phi(z)
   -\exp\!\left(\mu_E+\frac{\sigma_E^2}{2}\right)\Phi(z-\sigma_E),
 \end{aligned}
 \label{eq:appendix-ei}
\end{equation}
where $\phi$ and $\Phi$ are the standard normal density and cumulative distribution function. The zero-variance limit is $\max(E^\star-\exp(\mu_E),0)$. Multiplying this expectation by the predicted SLO and execution probabilities and dividing by predicted GPU-hours yields the acquisition in the main text. Before a feasible incumbent exists, selection uses feasibility per unit cost as specified in Section~\ref{sec:eco-acquisition}.

% Paragraph role: design; explain how prior updates preserve the residual target.
When calibration records change the stage model, ECO recomputes the prior prediction and residual target for every successful observation before refitting the GP. Each target is $(\log E_i-m_0(x_i))/s_0(x_i)$, as defined in Section~\ref{sec:eco-surrogates}, so the GP learns the error of the updated prior. Completed SLO violations supply residual targets; execution failures do not.

\subsection{Measurement Allocation and Screening Defaults}
\label{app:search-policy}

Following Section~\ref{sec:eco-selection}, ECO interleaves acquisition-based choices with probes of untried structures and single-control probes after the reference measurement. A structure probe collects calibration records at one operating point of the new structure. A single-control probe changes one role's frequency or power cap relative to a measured reference in the same structure, supplying the paired observations used in Section~\ref{app:surrogate-details}. These probes update the corresponding structure's model within the search allowance; a structure probe alone does not identify its control-response coefficients. A power-cap trial requires a cap at most 95\% of the busiest GPU's measured 95th-percentile power, so the cap could affect execution. Repeats normally wait until the final quarter of the allowance, except near measured SLO boundaries or when no new candidate remains. All probes and repeats launched during search count toward the 16-attempt search allowance; the shared preparation measurement is accounted for separately as specified in Section~\ref{app:eligibility}.

At acquisition steps, optimistic predictions use $\mu-2.5\sigma$. Candidates are deferred if these predictions still violate an SLO or exceed the best measured feasible energy by more than 2\%. ECO periodically tests deferred candidates and reconsiders them after model updates or when the active pool is empty. If acquisition scores numerically underflow, selection uses energy uncertainty weighted by SLO feasibility. Configurations that violate static constraints remain excluded; final eligibility depends on measured outcomes.

\subsection{BO and GA Baseline Configurations}
\label{app:baseline-settings}

\textbf{Shared protocol and initialization.}
Table~\ref{tab:baseline-hyperparameters} specifies the baseline settings used in each independent search run. Both methods search the scenario's statically legal candidate set, use the same calibration workload and fixed SLO thresholds, and maintain independent observation histories. Each begins by measuring the designated reference configuration as its first search attempt. This measurement counts toward the 16 attempts, separately from the preparation charge in Section~\ref{app:eligibility}. BO then starts acquisition-based selection, without an additional random initialization batch. GA fills a four-candidate initial population with the reference and three uniformly sampled, previously untried affordable candidates. Thus a complete GA run spends four attempts on initialization and twelve on offspring evaluations. Failed initial trials still consume attempts. For each scenario, the reported comparison averages results over multiple random seeds for each method, as specified in Section~\ref{sec:experiment-setup}. Each seed defines an independent search run with its own 16-attempt allowance and frozen configuration for held-out evaluation.

\begin{table}[htbp]
\centering
\small
\caption{BO and GA settings for comparisons averaged over multiple random seeds. Budgets and initialization apply to each independent search run. The population size in GA is the number of retained candidates, not an additional evaluation budget.}
\label{tab:baseline-hyperparameters}
\begin{tabularx}{\linewidth}{@{}lXX@{}}
\toprule
Setting & Constrained BO & GA \\
\midrule
Initialization & Reference, then acquisition & Reference + 3 random candidates \\
Search allowance per run & 16 attempts & 16 attempts \\
Search seeds & Multiple random seeds & Multiple random seeds \\
Maximum attempts per candidate & 2 & 1 \\
Energy prior & Constant $\log E_{\mathrm{ref}}$, SD 1 & No surrogate \\
Constraint prior & Mean 0, SD 0.5 & Measured constraint domination \\
GP kernel & Mat\'ern-$5/2$, length scale 1 & --- \\
GP observation noise & Log-scale SD 0.05 & --- \\
GP hyperparameter fitting & Disabled & --- \\
Population / tournament size & --- & 4 / 2 \\
Crossover probability & --- & 0.9 per offspring \\
Mutation probability & --- & 0.2 per gene \\
Mechanism prior / screening & Disabled / disabled & Disabled / disabled \\
\bottomrule
\end{tabularx}
\end{table}

\textbf{Constrained BO: representation and fitting.}
The input concatenates a one-hot encoding of deployment structure with the four log frequency/cap values, each min--max normalized over the candidate set. Mechanism-derived features are disabled. Energy uses a constant prior mean $\log E_{\mathrm{ref}}$ and standard deviation 1, where $E_{\mathrm{ref}}$ is the historical Default energy. Each normalized log-SLO constraint in Equation~\ref{eq:eco-constraints} has prior mean 0 and standard deviation 0.5. Independent GPs use a unit-amplitude Mat\'ern-$5/2$ kernel with fixed length scale 1, without hyperparameter optimization or target-mean normalization. With prior standard deviation $s$, the standardized target is $(y-m)/s$ and the diagonal noise variance is $(0.05/s)^2+10^{-8}$. GPs are refitted after each completed observation; successful SLO-violating runs contribute energy and constraint targets. Failed runs contribute execution and cost information but no fabricated energy or latency target.

\textbf{Constrained BO: acquisition and repeats.}
BO enumerates eligible candidates and maximizes the energy-unit lognormal expected improvement in Appendix~\ref{app:ei}, multiplied by SLO and execution probabilities and divided by predicted GPU-hours. Before finding a feasible observation, it uses feasibility probability per predicted GPU-hour. SLO probability is the product of the independent constraint probabilities. Execution probability is $(1+\sum_{i\in S}w_i)/(2+\sum_{i\in S\cup F}w_i)$, where $S,F$ denote successful and failed trials and $w_i=\exp(-\|\phi(x)-\phi(x_i)\|_2^2)$ for the encoded input $\phi$. Mechanism screening and ECO's structure/control probe schedules are disabled. Candidates marked runtime-incompatible are excluded, and each candidate permits at most two attempts, including failures. Acquisition ties prefer fewer previous attempts and then the lexicographically larger candidate identifier. If all scores are at most $10^{-300}$, selection maximizes energy uncertainty times SLO probability, again breaking ties by candidate identifier.

\textbf{Trial-cost prediction.}
Both implementations first exclude candidates whose predicted cost exceeds the remaining GPU-hour budget. Their base estimate is $c_0(x)=[t_{\mathrm{eval}}G_x+d(x,x_{\mathrm{prev}})\max(G_x,G_{\mathrm{prev}})]/3600$, where $G_x$ is the active GPU count and $t_{\mathrm{eval}}$ is the scenario-specific \texttt{evaluation\_seconds} stored in the archived configuration. The switching allowance $d$ is 120 seconds for either a structure or control change and zero for an unchanged configuration; the first launch uses 120 seconds and $G_{\mathrm{prev}}=G_x$. BO corrects $c_0$ with a GP on measured log cost ratios, using prior standard deviation 0.5 and observation-noise standard deviation 0.25; its predicted log correction is clipped to $[-3,3]$. GA uses $c_0$ for affordability without fitting a cost GP or dividing fitness by cost. Actual GPU-hours include all lifecycle costs and failures.

\textbf{GA: representation and fitness.}
GA is a steady-state search that evaluates one offspring at a time. Its seven genes are topology, the execution-mode/microbatch pair, the decode/prefill DBO-threshold pair, and the four operating controls. Topology and each paired setting are atomic genes. After initialization, the population is reconstructed as the best four previously evaluated candidates under a lexicographic fitness rule. Feasible candidates rank first by energy; completed infeasible candidates rank next by total normalized violation, with energy breaking ties; failed or invalid measurements rank last. The violation is
\begin{equation}
 V(x)=\sum_{j\in\mathcal L}\max\!\left(\frac{L_j(x)}{\tau_j}-1,0\right)
       +\max\!\left(\frac{R_{\min}}{R(x)}-1,0\right),
\end{equation}
where $\mathcal L$ contains the scenario's latency constraints. Candidate identifiers break remaining fitness ties. Fitness uses only this GA arm's measured outcomes.

\textbf{GA: selection, variation, and repair.}
Each parent is selected by a size-two tournament sampled without replacement; the second tournament excludes the first parent. With probability 0.9, uniform crossover independently chooses each gene from either parent with equal probability; otherwise the child copies the first parent. Each gene then mutates independently with probability 0.2 to a uniformly chosen alternative value present in the legal space. The child is repaired to an untried affordable legal candidate with minimum gene-wise Hamming distance, with uniform random tie-breaking. Repair uses configuration distance, not predicted fitness of unmeasured candidates. Each step reinitializes Python's random generator with the string \texttt{ga-v1:0:}$t$, where $t$ is the number of previous search attempts. Previously attempted candidates, including failures, are excluded from future proposals. Both methods stop when the attempt budget is exhausted or no admissible affordable candidate remains, and return the lowest-energy measured feasible configuration under Section~\ref{app:eligibility}.

The table reports settings verified from the available archived configuration manifests.

\section{Supplementary Search Statistics and Configurations}
\label{app:supplementary-search}

\subsection{Statistics for the Search Trajectories}

% Paragraph role: evidence; expand the search trajectories in the main text.
Table~\ref{tab:appendix-search} summarizes the DeepSeek configuration-search trajectory in Figure~\ref{fig:search-trials} and supplementary Qwen search records. Attempt numbers are one-based and include execution failures. ``Feasible'' counts completed attempts satisfying the full SLO rule, including Qwen's TTBT constraint. The selected attempt minimizes energy among feasible observations in each recorded trajectory. These quantities describe configuration-search progress; held-out measurements do not enter the trajectory or alter its selected attempt.

% Generated by data/appendix/build_tables.py; edit the generator.
\begin{table}[htbp]
\caption{Configuration-search statistics: DeepSeek in Figure~\ref{fig:search-trials} and supplementary Qwen records. Qwen uses eight A100 GPUs at 16 requests/s; DeepSeek uses four A6000 GPUs at 4 requests/s. Energy is for the selected search attempt.}
\label{tab:appendix-search}
\centering\small
\setlength{\tabcolsep}{4pt}
\begin{tabular}{@{}llrrrrr@{}}
\toprule
Scenario & Method & Attempts & Completed & Feasible & Selected & Energy (kJ) $\downarrow$ \\
\midrule
Qwen / A100 & ECO & 16 & 16 & 10 & 15 & 36.216 \\
Qwen / A100 & BO & 16 & 16 & 15 & 13 & 49.751 \\
DeepSeek / A6000 & ECO & 16 & 16 & 13 & 10 & 17.393 \\
DeepSeek / A6000 & BO & 16 & 16 & 11 & 14 & 26.503 \\
DeepSeek / A6000 & GA & 16 & 13 & 9 & 9 & 30.931 \\
\bottomrule
\end{tabular}
\end{table}

% Paragraph role: definition and evidence; make the common-target calculation reproducible.
For a common energy target $E_{\mathrm{target}}$, we count attempts to target as
\begin{equation}
 k_{\mathrm{target}}=\min\{k:\text{attempt }k\text{ completes, satisfies all SLOs, and }E_k\leq E_{\mathrm{target}}\}.
 \label{eq:appendix-target}
\end{equation}
In DeepSeek's four-A6000, 4-request/s trajectory, GA's best feasible energy gives $E_{\mathrm{target}}=30.9306$\,kJ. ECO, BO, and GA first reach this target at attempts 4, 10, and 9, respectively. This statistic counts measured attempts, including earlier failures; it does not estimate elapsed search time. The best feasible energies and target crossings are computed from unrounded observations, with rounding applied only for presentation.

% Paragraph role: evidence and limitation; connect counts to the existing interpretation.
For Qwen, ECO selects a lower-energy configuration with 10 feasible attempts, compared with BO's 15. ECO's attempt 16 has lower energy than its incumbent but violates TTFT, leaving attempt 15 selected. DeepSeek's GA attempts 2, 4, and 11 fail execution. They count toward its 16-attempt budget and appear as gaps in numerical measurements. Attributing these search outcomes to individual ECO components requires matched ablations.

\subsection{Selected A6000 Deployment Settings}
\label{app:selected-configurations}

% Paragraph role: evidence; supply concrete settings for the qualitative main-text analysis.
Table~\ref{tab:appendix-configs} lists ECO's search-selected A6000 configurations for the scenarios in Figure~\ref{fig:a6000-comparison}. All listed configurations use $\mathrm{TP}_A=\mathrm{TP}_F=1$ and enable DBO, so $\mathrm{DP}_A=N_A$, $\mathrm{DP}_F=N_F$, and $\mathrm{EP}_F=N_F$. The table reports role counts rather than host-specific GPU indices. Settings remain fixed during each replay.

% Generated by data/appendix/build_tables.py; edit the generator.
\begin{table}[htbp]
\caption{ECO's selected A6000 settings. Frequency and power-cap pairs are attention/expert; DBO thresholds are decode/prefill token counts. $G$ is the available GPU budget, and load is in requests/s. Power caps are per GPU.}
\label{tab:appendix-configs}
\centering\small
\setlength{\tabcolsep}{4pt}
\begin{tabular}{@{}lrrlrrr@{}}
\toprule
Model & $G$ & Load & Allocation & Clock (MHz) & Cap (W) & DBO thresholds \\
\midrule
Qwen & 4 & 4 & 2A+2E & 1560/1560 & 250/250 & 64/1024 \\
Qwen & 4 & 8 & 2A+2E & 1560/1560 & 200/250 & 64/1024 \\
Qwen & 8 & 8 & 4A+2E & 1560/1560 & 300/300 & 32/512 \\
Qwen & 8 & 16 & 4A+2E & 2100/1920 & 300/300 & 32/512 \\
DeepSeek & 4 & 4 & 2A+1E & 1560/1560 & 250/250 & 64/1024 \\
DeepSeek & 4 & 8 & 2A+1E & 1560/1560 & 250/250 & 64/1024 \\
DeepSeek & 8 & 8 & 4A+1E & 1560/1560 & 200/200 & 64/1024 \\
DeepSeek & 8 & 16 & 5A+1E & 1560/1560 & 200/200 & 64/1024 \\
\bottomrule
\end{tabular}
\end{table}

% Paragraph role: analysis; connect a supported configuration change to the main result.
For Qwen under the eight-GPU budget, both loads select 4A+2E (Section~\ref{sec:results-a6000}). Increasing load from 8 to 16 requests/s raises attention frequency from 1560 to 2100\,MHz and expert frequency from 1560 to 1920\,MHz, while retaining the same power caps and DBO thresholds. DeepSeek selects 2A+1E under the four-GPU budget and asymmetric allocations under the eight-GPU budget. Because the search selects the deployment structure and operating controls jointly, these rows do not isolate the energy saving from any one configuration choice.

\section{Configuration Characterization Details}
\label{app:characterization}

% Paragraph role: provenance and scope; explain what was counted.
Figure~\ref{fig:afd-characterization} uses configuration-search measurements and frozen validator outputs. The candidate sets fix the GPU pool, model, operating-point grids, and DBO threshold profiles. Their enumerator already applies allocation and rank-product restrictions. We count statically legal candidates by excluding model-divisibility and proven memory-capacity violations. An execution-verified structure does not establish SLO feasibility for all of its operating points.

For Qwen on four A6000 GPUs, static checks reduce 6,075 enumerated candidates to 2,700. Eight-GPU enumeration yields 31,725 candidates per model, of which 15,525 remain statically legal for Qwen and 22,950 for DeepSeek. The corresponding 16-attempt coverage is 0.593\%, 0.103\%, and 0.070\%.

% Paragraph role: outcomes; retain the distinction between a space and its sampled subset.
Each audited campaign evaluates 16 distinct legal candidates in 16 search attempts and charges one preparation evaluation separately, for 17 charged evaluations in total. Candidate-space coverage uses the 16 search attempts, consistently with Sections~\ref{sec:problem-formulation} and~\ref{sec:experiment-setup}. The leftmost panel of Figure~\ref{fig:afd-characterization} reports coverage by these attempts. Qwen on four A6000 GPUs at 8 requests/s completes 14 attempts, of which 9 satisfy the SLOs. For Qwen on eight GPUs, the completion/feasible counts are 16/4 at 8 requests/s and 15/2 at 16 requests/s. DeepSeek on eight GPUs completes all 16 attempts at each load, with 13 feasible at 8 requests/s and 12 at 16 requests/s. These counts describe the measured subsets of legal spaces containing 2,700, 15,525, and 22,950 candidates, respectively. The unmeasured candidates' execution and SLO outcomes are unknown.

% Paragraph role: interaction definition and evidence; avoid an independence claim from one point.
For two factors $a,b$, with all other settings and the replay workload fixed, define the energy interaction contrast
\begin{equation}
 I(a,b)=E(a,b)-E(a,b_0)-E(a_0,b)+E(a_0,b_0).
 \label{eq:afd-interaction}
\end{equation}
An additive energy response has $I=0$; a measured nonzero value also contains measurement noise. We inspect all fully observed structure-by-control rectangles within the five audited campaigns and find one in the current records. For the heatmap in Figure~\ref{fig:afd-characterization}, $a_0=(32,512)$ and $a=(64,1024)$ are DBO threshold pairs, while $b_0=2010$ and $b=2100$\,MHz are FFN frequencies. The energy difference caused by the frequency change is $+569.98$\,J at $a_0$ and $-732.60$\,J at $a$, yielding $I=-1302.58$\,J, or $-2.414\%$ of $E(a_0,b_0)$. The quartet shares the same search-workload trace, context, and execution protocol, and has one observation per cell. It exhibits a frequency-preference reversal, while the energy ranking between the two DBO settings remains unchanged. Three cells fail the throughput SLO. No confidence interval or general interaction-strength ranking follows from this quartet.

The quartet uses Qwen at 16 requests/s with 4A+2F, $\mathrm{TP}_A=\mathrm{TP}_F=1$, a 2100\,MHz Attention clock target, and 300\,W power caps. Only the DBO thresholds and FFN clock target vary.

% Paragraph role: scope of the recorded interaction evidence.
The quartet records end-to-end metrics but no stage timing or waiting measurements, so it cannot locate a migrating bottleneck.

\section{Motivation Microbenchmark Details}
\label{app:microbenchmarks}

\begin{figure}[t]
  \vspace{-6pt}
  \centering
  \setlength{\abovecaptionskip}{3pt}
  \begin{minipage}[t]{0.9\linewidth}
    \vspace{0pt}
    \input{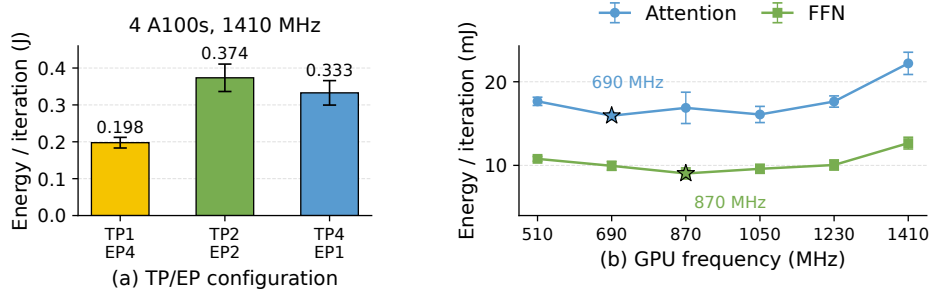}
  \end{minipage}
  \caption{Qwen-shaped A100 microbenchmarks: (a) TP/EP energy at 1410\,MHz; (b) Attention and FFN energy across frequencies.}
  \label{fig:motivation-study}
  \label{fig:joint-tuning-evidence}
  \vspace{-10pt}
\end{figure}

% Paragraph role: scope of the measured microbenchmarks.
Figure~\ref{fig:motivation-study} uses Qwen-shaped microbenchmarks on A100 GPUs. The recorded hidden and intermediate dimensions are 2048 and 512. The configuration comparison and operator frequency sweep probe different sources of energy variation; they are separate from the end-to-end serving workload in Section~\ref{sec:experiments}.

% Measured topology and frequency microbenchmarks.

% Paragraph role: design; explain the fixed factors in the topology experiment.
The parallel-configuration experiment fixes four GPUs, 6144 global tokens, one microbatch, and a 1410\,MHz clock. It compares TP1$\times$EP4, TP2$\times$EP2, and TP4$\times$EP1 with five repetitions per configuration. The TP/EP labels describe the partitioning in this microbenchmark; serving configurations use the vLLM parallelism convention in Table~\ref{tab:design-space}. The measured mean energies are 0.1976, 0.3735, and 0.3327\,J per iteration, with sample standard deviations of 0.0144, 0.0372, and 0.0332\,J, respectively.

% Paragraph role: design; specify the operator workload and sweep domain.
The frequency experiment measures attention and FFN prefill operators separately, using batch size one and 128 tokens. Each operator is evaluated at 510, 690, 870, 1050, 1230, and 1410\,MHz, with five repetitions per point. Requested and applied clocks agree in the recorded measurements. This sweep explores operator response over a broader frequency range than the serving search grid in Table~\ref{tab:design-space}; its frequencies do not extend that candidate grid.

\clearpage
% Generated by data/appendix/build_tables.py; edit the generator.
\begin{table}[htbp]
\caption{Operator frequency sweep underlying Figure~\ref{fig:motivation-study}(c). Energy is in mJ per iteration; entries are mean $\pm$ one sample standard deviation over five runs. Lower energy is better.}
\label{tab:appendix-microbenchmarks}
\centering\small
\setlength{\tabcolsep}{4pt}
\begin{tabular}{@{}rrr@{}}
\toprule
Frequency (MHz) & Attention (mJ) $\downarrow$ & FFN (mJ) $\downarrow$ \\
\midrule
510 & $17.66\pm0.49$ & $10.78\pm0.29$ \\
690 & $15.94\pm0.29$ & $9.95\pm0.54$ \\
870 & $16.88\pm1.87$ & $9.03\pm0.66$ \\
1050 & $16.09\pm0.97$ & $9.59\pm0.54$ \\
1230 & $17.64\pm0.67$ & $10.05\pm0.61$ \\
1410 & $22.20\pm1.33$ & $12.66\pm0.70$ \\
\bottomrule
\end{tabular}
\end{table}

% Paragraph role: definition and limitation; state exactly what the error bars support.
For five energies $e_1,\ldots,e_5$, the plotted mean and dispersion are $\bar e=\sum_j e_j/5$ and $s=\sqrt{\sum_j(e_j-\bar e)^2/4}$. Error bars show one sample standard deviation. Stars identify the lowest observed means, at 690\,MHz for attention and 870\,MHz for FFN. These minima describe the sampled operator workloads and do not establish statistically distinct or universal optima. End-to-end configuration selection additionally evaluates communication, waiting, and the service constraints defined in the main text.

\end{document}